\documentclass{mn2e}
\usepackage{times}
\usepackage{epsf}

\newcommand{\etal}{{et al}\/.}
\newcommand{\ngc}{\hbox{NGC\,315}}
\newcommand{\asca}{\textit{ASCA}}
\newcommand{\chandra}{\textit{Chandra}}
\newcommand{\rosat}{\textit{ROSAT}}

\newcommand{\hst}{\textit{HST}}

\begin{document}

\title[The X-ray jet and central structure of \ngc]{The X-ray jet and central structure of
the active galaxy \ngc}

\author[D.M. Worrall, M. Birkinshaw \& M.J. Hardcastle]{D.M. Worrall,
M. Birkinshaw \& M.J. Hardcastle\\
Department of Physics, University of Bristol, Tyndall Avenue,
Bristol BS8~1TL}

\maketitle

\label{firstpage}

\begin{abstract}
We report the \chandra\ detection of resolved X-ray emission
of luminosity $3.5 \times 10^{40}$ ergs s$^{-1}$ (0.4--4.5 keV)
and power-law energy spectral index $\alpha = 1.5 \pm 0.7$ from a roughly 10 arcsec
length of the north-west radio jet in \ngc.  The X-ray emission is
brightest at the base of the radio-bright region about 3 arcsec from
the nucleus, and is consistent with a synchrotron origin.
At a projected distance of 10 arcsec from the core, the jet is in
approximate pressure balance with an external medium which is also
detected through its X-ray emission and which has $kT \approx 0.6 \pm
0.1$~keV, consistent with earlier \rosat\ results.  The high spatial
resolution and sensitivity of \chandra\ separates nuclear unresolved emission
from the extended thermal emission of the galaxy atmosphere with
higher precision
than possible with previous telescopes.
We measure an X-ray luminosity of
$5.3 \times 10^{41}$ ergs s$^{-1}$ (0.4--4.5 keV) and a power-law
energy index of $\alpha = 0.4 \pm 0.4$ for the nuclear component.
\end{abstract}


\section{Introduction}
\label{sec:intro}

X-ray studies of the kpc-scale radio jets of active galaxies can test
whether their electrons are accelerated to high energies.  Fast
synchrotron energy losses of high-energy electrons allow X-ray maps
to pinpoint the acceleration sites.  Nearby, low-power jets of the
Fanaroff \& Riley (1974) class I (FRI) have large projected angular
sizes, and provide the best opportunity for measuring X-ray
substructure.

The high spatial resolution and sensitivity of \chandra\ have been
crucial to extending X-ray jet studies beyond Cen\,A and M\,87, the
only FRI radio galaxies for which resolved X-ray jets had been
detected prior to its launch (Schreier et al.~1979; Schreier,
Gorenstein \& Feigelson 1982).  In this paper we report the detection
of an X-ray jet in the inner parts of the giant FRI radio galaxy
NGC~315, at $z=0.0165$.

The observation was made as part of our \chandra\ programme of imaging
and spectroscopy of FRI radio galaxies from the B2 bright sample
(Colla et al. 1975; Ulrich 1989), aimed at studying the validity of
the unification of B2 radio galaxies and BL Lac objects and at testing
physical and dynamical models for the X-ray and radio-emitting
structures. \ngc\ is one of the X-ray brightest B2 sources studied
with \rosat\ (Canosa et al.~1999), and consequently only a short
\chandra\ exposure time was needed to meet our criterion of uniform
data quality for the sample.

NGC\,315 has a two-sided radio structure extending roughly one degree
on the sky (Bridle et al.~1979).  On milliarcsec scales, the sidedness
asymmetry and proper motions suggest the jet bulk motion is
relativistic ($v > 0.75 c$), possibly accelerating between 3 and 10 pc
from the nucleus, with an angle to the line of sight between 30 and 40
degrees (Cotton et al.~1999; Giovannini et al. 1994, 2001).  The north-west radio
jet remains dominant on a scale of a few tens of arcsec.  A transition
from supersonic to turbulent transonic flow, with deceleration by
entrainment, is likely to occur at about 17 arcsec from the core,
where there is a sudden change in jet opening angle (Bicknell 1986).
Bicknell (1994) shows from the application of conservation laws that
an entraining jet which is initially relativistic will have a speed
between $0.3$ and $0.7~c$ at the flare point, and he demonstrates how
such a model is consistent with the radio data from \ngc.

Resolved jet emission from \ngc\ remains unseen in the optical from
ground-based (Butcher et al.~1980) and \hst\ (Verdoes Kleijn et
al.~1999) measurements.  The \hst\ observations detect an unresolved
nuclear source surrounded by an inclined, regular,
circum-nuclear disk of radius about 1 arcsec and extinction about 0.25
magnitudes, whose axis is aligned to within a few degrees of the north-west
radio jet (Verdoes Kleijn et al.~1999; de Ruiter et al. 2002).

In this paper we adopt a value for the Hubble constant of $H_o =
70$~km s$^{-1}$ Mpc$^{-1}$.  1~arcsec corresponds to 335~pc at \ngc.  The
J2000 position of the radio nucleus is $\alpha = 00^\circ 57'
48''.8834$. $\delta = +30^{\rm h} 21^{\rm m} 08^{\rm s}.812$ (Xu
\etal\ 2000). The Galactic column density to the source obtained from
21~cm measurements is $5.92 \times 10^{20}$ cm$^{-2}$ (Dickey \&
Lockman 1990), and local absorption through this gas is included in
all our spectral models.

\section{\chandra\ Observations and Results}
\label{sec:obs}

We observed \ngc\ with the back-illuminated CCD chip, S3, of the
Advanced CCD Imaging Spectrometer (ACIS) on board \chandra\ on 2000
October 8.  Details of the instrument and its modes of operation can
be found in the \chandra\ Proposers' Observatory Guide, available from
http://cxc.harvard.edu/proposer. The observation was made in the
VFAINT data mode, with a 128-row subarray (giving a 1 by 8 arcmin
field of view, with the source roughly centred in the 1 arcmin
direction and at a 3:1 offset position in the 8 arcmin direction).
The use of the subarray was to reduce the readout time to 0.44~s in
order to guard against the effects of pile-up should much of the flux
detected with {\it ROSAT\/} (Worrall \& Birkinshaw 1994) be unresolved
to {\it Chandra\/}. Results presented here all use {\sc ciao v2.3} and the {\sc
caldb v2.20} calibration database.  The data have been re-calibrated
and analysed, with random pixelization removed and taking advantage of
VFAINT background cleaning, following the software `threads' from the
\chandra\ X-ray Center (http://cxc.harvard.edu/ciao).  Events with
grades 0,2,3,4,6 are used.  All X-ray spectra are binned to a minimum
of 20 counts per bin, and the time-dependent decline in quantum
efficiency is taken into account using the recommended methods.  After
the VFAINT background cleaning, there were no time intervals which
needed to be excluded because of anomalously high background, and the
calibrated data have an observation duration of 4.67~ks.

\subsection{The X-ray image and resolved jet}
\label{sec:ximage}

\begin{figure}
\epsfxsize 7.0cm
\epsfbox{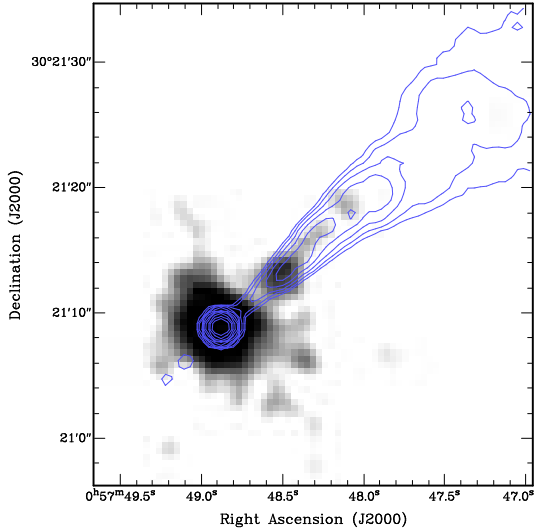}
\caption{
The grey-scale image shows the 0.5--5 keV \chandra\ data for \ngc, in
$0.492''$ bins and smoothed with a Gaussian of  $\sigma = 0.74''$.
The contours (in factors of two increment) outline the radio jet
from a high-resolution 5~GHz VLA image (Cotton, 2002,
private communication)
which has been convolved with a Gaussian of $\sigma = 1''$ in
order to match the resolution of the X-ray image.
Jet X-ray emission is seen, particularly at the base of the first radio
enhancement. Low surface-brightness X-ray emission is suppressed in
this image. 
} 
\label{fig:image}
\end{figure}

Fig.~\ref{fig:image} shows the 0.5--5 keV \chandra\ X-ray data
together with contours outlining the inner radio emission. In the
figure, the X-ray core is burned out and the low-surface-brightness
emission is suppressed in order to emphasize small-scale extended
emission.

X-ray emission from the core is extended with a similar ellipticity
and position angle to the optical emission from \ngc.  We have used
the {\sc IRAF STSDAS} task {\sc ELLIPSE} to measure the ellipticity of
a Kitt Peak 4~m R-band image (Birkinshaw and Davies, private
communication), and find position angles between 40 and 46 degrees and
ellipticities between 0.16 and 0.26 for most of the galaxy.  For
example, at a semi-major axis of 10 arcsec, the ellipticity is $0.251
\pm 0.002$ and the position angle is $45.2 \pm 0.2$ degrees.  We find
an ellipticity of $0.29\pm 0.03$ and a position angle of $49\pm 3$
when similar procedures are applied to the 0.5-5~keV X-ray image smoothed with a
Gaussian of $\sigma = 1.3$ arcsec, masking out the region of the
resolved jet.

Resolved X-ray emission from the NW radio jet is clearly evident at a
position angle roughly perpendicular to the semi-major axis of the
galaxy. The X-ray emission is particularly strong at the base of
the first radio enhancement.  We find
$55 \pm 9$ net counts in the energy range $0.5-5$~keV from the jet as
measured in a rectangular box of length 12 arcsec and width 6 arcsec, starting
3 arcsec from the nucleus, taking a corresponding box diametrically
opposite the nucleus to measure the background.

\subsection{Radial-Profile Analysis}
\label{sec:radial}

In order to separate unresolved and galaxy emission, avoiding
resolved jet emission, 
we extracted a source-centred 0.4--7 keV radial profile
excluding a pie slice of position angle between $295^\circ$ and
$325^\circ$.  We fitted the profile
with models convolved with the Point
Response Function (PRF), using home-grown software whose algorithms are
described in Birkinshaw (1994).  Our procedure for
finding the appropriate energy-weighted PRF follows that described in
Worrall et al. (2001), except that we used a revised functional form
for the third additive component in order to give an improved description of
bright point sources (see http://cxc.harvard.edu/cal/Hrma/psf):
for the energy distribution of the counts in our \ngc\ observation
we find this third component to be proportional to $(1 + \theta/0.021)^{-2.1}$, 
for $\theta \leq 5$ arcmin.

The \rosat\ PSPC measured thermal X-ray emission associated with \ngc's
atmosphere out to a radius of 2 arcmin, and we expect this emission to
dominate over the background throughout part of the field of view of
our \chandra\ observation.  A rectangular box of size 1 arcmin by 2
arcmin as far from \ngc\ as possible but still in the field of view
(i.e. all being 4 arcmin or more away from the nucleus of \ngc)
contains $0.007 \pm 0.001$ cts arcsec$^{-2}$, 0.4--7 keV,
consistent with time-averaged results available from
http://cxc.harvard.edu/cal/Acis/Cal\_prods/bkgrnd/current.
We use this
as a measure of the background in the observation for all the analyses
presented here.  As a check, we have also performed the radial
analysis allowing our fitting to find the best-fitting source and
background contributions to an annulus of radii 40 and 62 arcsec, and
the results for the background and the model parameters
agree with those using the rectangular-box background.

\begin{figure}
\epsfxsize 7.0cm
\epsfbox{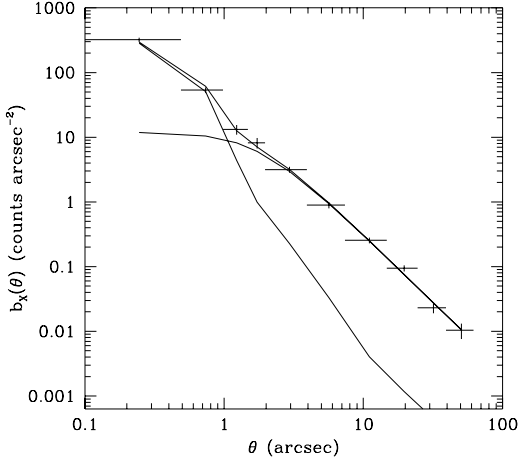}
\caption{
The data show the background-subtracted 0.4--7 keV radial profile,
excluding a pie slice of position angle between $295^\circ$ and
$325^\circ$ in order to avoid the resolved jet emission.  The model is a
composite of a $\beta$ model and unresolved component: the individual
components and the sum are shown as solid lines.  The counts per unit
area in a $\beta$ model are proportional to $(1 + \theta^2/\theta_{\rm
cx}^2)^{0.5 - 3\beta}$, and the best fit values are $\beta = 0.515$,
$\theta_{\rm cx} = 1.55$ arcsec. $\chi^2_{\rm min} = 8.9$ for 6
degrees of freedom.
} 
\label{fig:profile}
\end{figure}

The radial profile gives unacceptable fits to single-component models
of a point source or a $\beta$ model (used to describe gas in
hydrostatic equilibrium), but the fit to the combination of the two is
good, as shown in Fig.~\ref{fig:profile}.  Parameter values and
uncertainties are given in Table~\ref{tab:radial}.

\begin{table}
\caption{Radial-profile analysis, 0.4--7 keV}
\label{tab:radial}
\begin{tabular}{ll}
Parameter & Value \\
point-source counts & $380 \pm 35$ \\
$\beta$ & $0.515 \pm 0.03$ \\
Core radius, $\theta_{\rm cx}$ & $1.55 \pm 0.7$ arcsec \\
$\theta_{\rm FW10\%M}$ & $9 \pm 3$ arcsec, $3.0 \pm 1$ kpc \\
$\beta$-model counts, $\theta < 10$ arcsec & $409 \pm 40$ \\
$\beta$-model counts, $\theta < 45$ arcsec & $693 \pm 44$ \\
\end{tabular}
\medskip
\begin{minipage}{\linewidth}
Since parameters are correlated ($\beta$, $\theta_{\rm cx}$), (point-source
counts, $\beta$-model counts), uncertainties are $1\sigma$ for two
interesting parameters.  $\theta_{\rm FW10\%M}$ is the full width to
10 per cent of maximum of the $\beta$ model.  Counts have been multiplied by
360/330 to account for the excluded pie slice.
\end{minipage}
\end{table}

\subsection{X-ray Spectrum}
\label{sec:spectrum}

\begin{figure}
\epsfxsize 5.0cm
\epsfbox{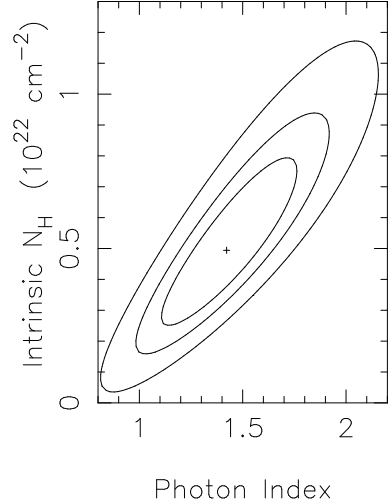}
\caption{
$\chi^2$ contours ($1\sigma$, 90\%, and 99\%, for two interesting
parameters) for a spectral fit to the nuclear X-ray emission extracted
from a circle of radius 1 arcsec.  The model is a single-component power
law with intrinsic and (fixed) Galactic absorption.  In the band 0.4 -
7 keV the spectrum contains 313 counts, and from the radial profile we
expect roughly 10 per cent of these counts to be from the extended hot-gas
component.  $\chi^2_{\rm min} = 11.5$ for 13 degrees of freedom.
} 
\label{fig:corecont}
\end{figure}

Our radial-profile analysis separates the 313 counts\footnote{Counts
refer to 0.4--7~keV throughout \S 2.3.} within a source-centered
circle of radius 1 arcsec into contributions of 90 per cent from
unresolved emission and 10 per cent from the extended component.
Spectral analysis supports the dominance of non-thermal emission in
the unresolved component.  The data give a good fit to a
single-component absorbed power law, as shown in
Fig.~\ref{fig:corecont}.  While the statistics are inadequate for
fitting a two-component model, a single-component thermal model gives
unrealistic results: a best-fitting temperature of 25 keV, an
intrinsic absorption of $N_{\rm H} = 4 \times 10^{21}$ cm$^{-2}$, and
a 0.5 -- 7 keV luminosity of $3 \times 10^{41}$ ergs s$^{-1}$ which is
unreasonably high for the X-ray binary population within 335~pc of
\ngc's centre.

To measure the spectrum of the extended emission we have fitted a
Raymond-Smith thermal model to the 511 counts from a core-centred annulus
of radii 2 arcsec and 45 arcsec, excluding a pie slice of position
angle between $295^\circ$ and $325^\circ$, to avoid the resolved jet
emission and the majority of counts from the unresolved emission.  The
best fit gives $kT \approx 0.6$~keV and a small intrinsic absorption
of $N_{\rm H} = 5 \times 10^{20}$ cm$^{-2}$ (comparable to Galactic
absorption), although the uncertainties are such that the additional
absorption is consistent with being zero.  There is marginal
improvement in the fit (an $F$ test giving a 7 per cent probability of the
improvement in $\chi^2$ being by chance) if a thermal bremsstrahlung
component (best-fitting $kT = 4.2$ keV) is included in the model.  In this
case the minimum $\chi^2$ is 20.8 for 17 degrees of freedom, and a
0.5--7 keV luminosity of $6 \times 10^{40}$ ergs s$^{-1}$ (about 20
per cent of the total luminosity from the region) is in the bremsstrahlung
component.  Although this is a factor of a few higher than
expected for the composite emission from X-ray binaries in the
\ngc\ galaxy, the uncertainties are large.  The abundance of the
thermal gas increases from about 0.2 solar to more reasonable values
between about 0.5 and 1.0 solar when the bremsstrahlung component is
included.  A consistent temperature is found for the thermal gas,
whether or not the bremsstrahlung component is included, and the
statistical uncertainties dominate the errors.  The best-fit value and
90 per cent confidence uncertainties (for one interesting parameter) are
estimated to be $kT = 0.62\pm 0.06$ keV.

\begin{figure}
\epsfxsize 5.0cm
\epsfbox{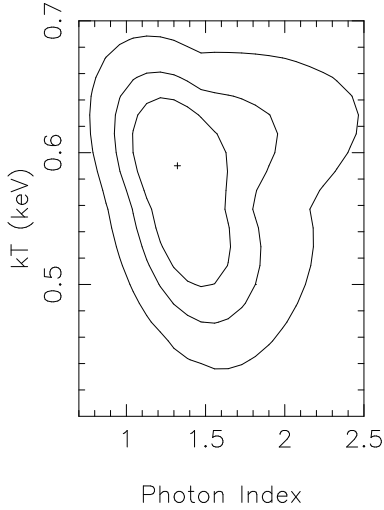}
\caption{
$\chi^2$ contours ($1\sigma$, 90\%, and 99\%, for two interesting
parameters) for a spectral fit to the X-ray emission from a circle of
radius 45 arcsec, excluding a pie slice of position angle between
$295^\circ$ and $325^\circ$, to avoid the resolved jet emission The
model is the combination of a power law with intrinsic absorption and
a thermal model, both with (fixed) Galactic absorption.  $\chi^2_{\rm
min} = 36.1$ for 33 degrees of freedom.
} 
\label{fig:both2cont}
\end{figure}

Our spectral fitting to the total 920 counts in the central 2 arcsec radius
core plus the region of the previous paragraph finds a two-component
power-law plus thermal model to be acceptable.  The statistical
quality of the data is insufficient to require more complex spectral
models, such as the addition of a bremsstrahlung component.  The
results are insensitive to the gas abundance (which we set to be 0.5
times solar) and intrinsic absorption in front of the thermal emission
(which we set to be zero).  The $\chi^2$ contours of power-law slope
versus gas temperature are shown in Fig.~\ref{fig:both2cont}.  For
the best-fit parameters, the number of counts in the power-law
component is $438 \pm 38$ ($1\sigma$ statistical error only).  This is in
reasonable agreement with the results from the radial-profile analysis
(Table~\ref{tab:radial}).  In the
remainder of this paper the radial-profile analysis is used to
separate the counts into unresolved emission, which we characterize by
a power law of photon spectral index $1.4 \pm 0.4$, from the resolved
emission which we characterize by a thermal model with $kT = 0.6\pm
0.1$ keV.

With so few counts, the spectrum of the resolved jet emission is
poorly constrained.  However, the distribution of counts suggests a
spectrum which is steeper than that of the core. A formal fit
(of the counts in four energy bins) with no intrinsic absorption
gives a photon spectral index of $2.5 \pm 0.7$ ($1\sigma$ uncertainty).

\subsection{Physical Parameters of the X-ray Components}
\label{sec:denspressure}

In this short observation we have measured resolved jet emission between about
3 and 15 arcsec from the nucleus.  The counts are converted into a
0.4-5 keV luminosity in Table~\ref{tab:physxpars} assuming a power law
of photon spectral index 2.5.  The 1~keV flux density is then 4.3~nJy.
The 5 GHz radio flux from a matched region in the map of Fig.~1 is 68
mJy. The two-point rest-frame 5-GHz to 1-keV spectral index is therefore $\alpha_{\rm rx}
= 0.94\pm 0.01$, where the error is dominated by the 16 per cent
uncertainty in the X-ray counts (\S 2.1).

\begin{table}
\caption{X-ray components}
\label{tab:physxpars}
\begin{tabular}{ll}
Parameter & Value \\
Resolved Jet: \\
$L_{\rm x}$ (0.4--5 keV) &  $3.5 \times 10^{40}$ ergs s$^{-1}$\\
Core: \\
$L_{\rm x}$ (0.4--5 keV) &  $5.3 \times 10^{41}$ ergs s$^{-1}$\\
Gas: \\
$L_{\rm x}$ (0.4--5 keV, $r \leq 15$ kpc) &
$3.6 \times 10^{41}$ ergs s$^{-1}$\\
Density $(\theta = 0)$ & $(3.4^{+1.5}_{-1.0})\ 10^{-1}$ cm$^{-3}$\\
Density $(\theta = 10'')$ & $(1.9 \pm 0.4)\ 10^{-2}$ cm$^{-3}$\\
Pressure $(\theta = 0)$ & $(7^{+3}_{-2})\ 10^{-11}$ Pa\\
Pressure $(\theta = 10'')$ & $(4 \pm 1)\ 10^{-12}$ Pa\\
$\dot{M} (\theta \leq \theta_{\rm cx})$ & 
$0.05 \pm 0.01$ M$_\odot$ yr$^{-1}$ \\
\end{tabular}
\medskip
\begin{minipage}{\linewidth}
Luminosities are before absorption.
Multiply pressure values by 10 for units of dynes cm$^{-2}$,
commonly quoted in X-ray astronomy papers.
Uncertainties are $1\sigma$ for 1 interesting parameter.
$\dot{M}$ is the mass cooling rate of gas within the core radius.
Uncertainties in the luminosities are dominated by the statistical
uncertainties in the counts, given earlier in the paper.
\end{minipage}
\end{table}

With regard to the core emission, the spectral index of the power-law
component (Fig.~\ref{fig:corecont}) is highly correlated with the
intrinsic absorption. Our results suggest a somewhat flatter spectral
index and lower absorption than found by Matsumoto et al.~(2001;
spectral index of $2.02^{+0.37}_{-0.32}$, $N_{\rm H} =
0.9^{+0.6}_{-0.5} \times 10^{22}$ cm$^{-2}$) in a two-component
thermal and power-law fit to \asca\ data, but the results are
consistent with a similar correlation of spectral index and
intrinsic absorption.  Table~\ref{tab:physxpars} gives the best-fit
0.4--5 keV luminosity of the power-law component.  This is comparable
with the luminosities of two other low-power radio galaxies from the
B2 sample in which \chandra\ has found unresolved X-ray cores and
resolved jets, B2 0206+35 at $2.9 \times 10^{41}$ ergs s$^{-1}$ and B2
0755+37 at $1.9 \times 10^{42}$ ergs s$^{-1}$ (Worrall et al.~2001).
The 2--10 keV luminosity of \ngc's power-law component is $6.8 \times
10^{41}$ ergs s$^{-1}$, larger than the \asca\ value of $3.1 \times
10^{41}$ ergs s$^{-1}$ (Matsumoto et al.~2001) and possibly supporting
variability similar to that reported between \rosat\ PSPC and HRI
observations by Worrall \& Birkinshaw (2000).  The interpolation to a
0.2--1.9 keV luminosity further supports variability of a factor of
about 3 when compared with the \rosat\ PSPC results (Worrall \&
Birkinshaw 2000), with \rosat\ being higher, but this is highly
dependent on the spectral index used for the interpolation, and a
spectrum closer to the \asca\ result removes the requirement for
variability.  The 2~keV core flux density from the \chandra\ observation
is 40~nJy.  Since this is the same value as adopted by Giovannini et
al.~(1994), their conclusions concerning relativistic beaming based on 
synchrotron self-Compton emission in the VLBI core would be unchanged.

Values for physical parameters of the X-ray-emitting gas are listed in
Table~\ref{tab:physxpars}.  \chandra's high spatial resolution finds
a smaller core radius (Table~\ref{tab:radial}) and a density gradient
within 10~arcsec of the centre to which the \rosat\ PSPC was
insensitive, but the gas pressure at 10 arcsec is in remarkable
agreement with the \rosat\ results (compare Table~\ref{tab:physxpars} with figure 9 of
Worrall \& Birkinshaw 2000).   When the best-fit \chandra\ results are
extrapolated to
estimate the gas luminosity for $\theta < 3$ arcmin and 0.2--2.5 keV,
we find a value of $5.3 \times 10^{41}$ ergs s$^{-1}$, which
is in excellent agreement with the value of $(5.4 \pm 0.2) \times
10^{41}$ ergs s$^{-1}$ from \rosat, and the gas temperature (see
Fig.~\ref{fig:both2cont}) is also in good agreement with the value
of $kT = 0.62 ^{+0.09}_{-0.1}$~keV from \rosat\ (Worrall \&
Birkinshaw 2000).
Our results for the temperature are lower than 
$kT = 0.8^{+0.04}_{-0.06}$~keV found from \asca\ data
by Matsumoto et al. (2001).  While there is no obvious explanation for
the difference, we note that while the \asca\ data have more counts,
the gas and AGN emission were both unresolved in the \asca\ beam so
that there was no ability to measure the spectrum of the gas
uncontaminated by AGN emission, as is possible with \chandra.
\ngc's 0.4--5 keV gas luminosity within a radius of 15 kpc
is similar to the values of $2.6 \times 10^{41}$ ergs s$^{-1}$ and
$4.2 \times 10^{41}$ ergs s$^{-1}$ found for B2 0206+35 and B2
0755+37, respectively
(Worrall et al.~2001). 

\section{Discussion}
\label{sec:discussion}

\chandra\ has 

\begin{enumerate}

\item made the first detection of resolved X-ray jet emission in the
nearby radio galaxy \ngc,

\item confirmed, with high spatial resolution, results from \rosat\
and \asca\ that the bulk of the X-ray emission is comprised of
a combination of galaxy-scale emission, principally from a hot X-ray-emitting
atmosphere, and unresolved emission which fits a power-law spectrum,
and

\item found evidence that the resolved X-ray jet is of steeper
spectrum than the core, and that the jet emission is brightest close to
the core, in particular at the base of the first bright radio knot.

\end{enumerate}

The resolved jet emission has properties similar to those seen in deep
observations of some other FRI sources.  For example, the power-law
energy spectral index of the X-ray emission, $\alpha_x = 1.5 \pm 0.7$,
is consistent with the two-point radio to X-ray spectral index of 
0.94, supporting a synchrotron origin with either a broken or
unbroken spectrum, depending on the radio spectrum in this region.
Such a synchrotron spectrum would be consistent with Butcher et al.~(1980)'s optical upper
limit on jet emission of $1\mu$Jy per kpc.
The inverse Compton arguments applied to 3C\,66B by Hardcastle et al.~(2001)
give similar results for \ngc\ in underpredicting the X-ray emission
under reasonable assumptions for orientation, beaming, and
available photon fields.  The strong X-ray emission at the base
of the first radio enhancement is a behaviour similar to that seen in
3C\,66B (Hardcastle et al.~2001) and Cen\,A (Kraft et al.~2002).

When we compare the X-ray flux density of the unresolved emission with
the total VLBI 5 GHz radio flux density of
$477 \pm 5$ mJy from Venturi et al.~(1993), we find a
two-point rest-frame 5~GHz to 1~keV spectral index of $\alpha_{\rm rx} \approx
0.86$.  The uncertainties are dominated by possible variability, with
the factor of three discussed in \S 2.4 leading
to an uncertainty in $\alpha_{\rm rx}$ of about $\pm 0.06$.
The value of  $\alpha_{\rm rx}$ is within 
the range of 0.8 to 0.93 found for three other B2 bright-sample
FRI radio galaxies observed with \chandra\ (Worrall et al.~2001).
Similar values of $\alpha_{\rm rx}$ for galaxies of different
fluxes would support a relationship between the core
radio and X-ray components, as argued statistically for the B2 bright
sample based on \rosat\ data by Canosa et al.~(1999).
The core two-point spectral index is somewhat
flatter than that inferred for the resolved jet, 
possibly lending support to a different (inverse Compton) origin for
the core emission (e.g., Giovannini et al.~1994, Hardcastle \& Worrall
2000).

The nuclear X-ray emission of \ngc\ is seen through a moderate
intrinsic column density (Fig.~3) of $N_{\rm H} =
(5^{+3.2}_{-2.7})\times 10^{21}$ cm$^{-2}$ (90 per cent confidence
uncertainty for 1 interesting parameter).  \chandra\ typically finds
intrinsic column densities below $10^{22}$ cm$^{-2}$ in FRI
radio-galaxy nuclei (e.g., Worrall et al.~2001; Hardcastle et
al.~2001, 2002; Wilson \& Yang 2002).  Nuclear optical emission, now
commonly detected with HST, has been shown by Chiaberge, Capetti \&
Celotti (1999) to correlate with the VLA-scale core radio emission.
This optical emission is generally associated with extinction at a
level compatible with the X-ray absorption (e.g., Ferrarese \& Ford
1999), suggesting that both the X-ray and optical emission are
associated with the sub-kpc-scale radio jets (Hardcastle \& Worrall
2000).

It is unclear whether FRI radio galaxies contain a dense torus of gas
and dust of the type required by models which unify powerful radio
sources (Barthel 1989).  Cen~A and NGC~4261 are the clearest cases
where column densities of order $10^{23}$ cm$^{-2}$ are required by
the X-ray data (Kraft et al.~2003, in preparation; Zezas et al.~2003).
Both objects are closer than NGC~315, with the advantage that the
nuclear emission is better separated from kpc-scale, radio-related,
X-rays.  However, M~87 is also nearby and does not require a
heavily-absorbed component (Wilson \& Yang 2002).  Even if central
tori exist in FRI radio galaxies, most of the X-ray emission may be
associated with structures on larger scales, since there is at least
one good case where the nucleus is known to be radiatively inefficient
(Di Matteo et al.~2003). Little or no excess nuclear absorbing column
would then be seen, except in cases (such as NGC~4261) where the jets
lie almost in the plane of the sky.  If this is the case, the low
X-ray column density of NGC 315, whose jets may lie $\sim 35^\circ$
from the line of sight (see \S 1), says little about the presence or
absence of a nuclear torus, although a second, heavily-absorbed,
component might appear in a deeper X-ray observations.

We have compared the pressure in the radio jet with that in the
external X-ray emitting gas, assuming an equipartition magnetic field.
At 10 arcsec from the core the FWHM of the radio jet is roughly 1.3
arcsec.  Using the radio flux density in a cylinder of length 1 arcsec,
modelling the electron spectrum above $\gamma_{\rm min} = 10$ with a
synchrotron spectrum consistent with the distribution of
radio, X-ray, and optical (upper limit)
flux densities over the surrounding 12-arcsec length of jet, assuming
no relativistic protons,
a filling factor of unity, no relativistic bulk motions, and a jet on
the plane of the sky, we find a pressure of about $6 \times
10^{-12}$~Pa, which is a reasonable match to the external pressure of
$(4\pm 1)\times 10^{-12}$ Pa (Table~\ref{tab:physxpars}) given the
measurement and model uncertainties.  Such a model would be
appropriate if the radio source is intrinsically one-sided.  If at 10
arcsec from the core the jet is inclined at $\theta = 35$ degrees to
the line of sight, and has a bulk velocity of $0.7 c$ (see \S 1), then
the bulk relativistic beaming factor, $\delta = 1/\gamma(1 -
\beta\cos\theta) = 1.67$.  The minimum pressure in the jet will now be
reduced, by a factor of roughly $\delta^{-10/7}(\sin\theta)^{4/7}$
(Bicknell 1994), i.e., to $2 \times 10^{-12}$~Pa.  Projection effects
mean that the jet is longer.  The pressure in a $\beta$-model external
atmosphere scales as $(\cos\theta)^{3\beta}$ at large distances from
the core radius, i.e., to give $\approx 2.9 \times 10^{-12}$ Pa, such
that a match between minimum jet pressure and external pressure can
still be claimed given the uncertainties.  Pressure balance between
the external medium and the minimum pressure in the jet at about 100
arcsec from the core (on the reasonable assumption that the jet is now
sufficiently slow that beaming can be ignored) has previously been
reported based on \rosat\ PSPC data by Worrall \& Birkinshaw (2000),
who point out that \ngc\ appears to be rather different from other B2
bright sample radio galaxies in remaining in pressure balance (at
minimum pressure) over a large length of radio jet, without requiring
an additional component of internal pressure.


\label{lastpage}

\end{document}